\documentclass[apl, reprint,superscriptaddress,showpacs,preprintnumbers,amsmath,amssymb, hidelinks, nofootinbib]{revtex4-1}

\pdfoutput=1

\usepackage{amsmath}
\usepackage{graphicx}
\usepackage{natbib}
\usepackage{verbatim}
\usepackage[withbib, all]{authorindex}	
\usepackage[usenames]{color}
\usepackage{bm}
\usepackage{hyperref}
\usepackage[export]{adjustbox}

\hypersetup{
    colorlinks, linkcolor={blue},
    citecolor={blue}, urlcolor={blue},
}

\begin{document}

\title{Spatially resolved detection of complex ferromagnetic dynamics using optically detected NV spins}

\author{C. S. Wolfe}
\affiliation{Department of Physics, The Ohio State University, Columbus, Ohio 43210, USA}

\author{S. A. Manuilov}
\affiliation{Department of Physics, The Ohio State University, Columbus, Ohio 43210, USA}

\author{C. M. Purser}
\affiliation{Department of Physics, The Ohio State University, Columbus, Ohio 43210, USA}

\author{R. Teeling-Smith}
\affiliation{Department of Physics, The Ohio State University, Columbus, Ohio 43210, USA}

\author{C. Dubs}
\affiliation{INNOVENT e.V. Technologieentwicklung, Pruessingstrasse 27 B, 07745 Jena, Germany.}

\author{P. C. Hammel}\email{hammel@physics.osu.edu*}
\affiliation{Department of Physics, The Ohio State University, Columbus, Ohio 43210, USA}

\author{V. P. Bhallamudi}\email{bhallamudi.1@osu.edu*}
\affiliation{Department of Physics, The Ohio State University, Columbus, Ohio 43210, USA}


\date{\today}

\begin{abstract}
We demonstrate optical detection of a broad spectrum of ferromagnetic excitations using nitrogen-vacancy (NV) centers in an ensemble of nanodiamonds. Our recently developed approach exploits a straightforward CW detection scheme using readily available diamond detectors, making it easily implementable. The NV center is a local detector, giving the technique spatial resolution, which here is defined by our laser spot, but in principle can be extended far into the nanoscale.  Among the excitations we observe are propagating dipolar and dipolar-exchange spinwaves, as well as dynamics associated with the multi-domain state of the ferromagnet at low fields.  These results offer an approach, distinct from commonly used ODMR techniques, for spatially resolved spectroscopic study of magnetization dynamics at the nanoscale. 
\end{abstract}

\pacs{07.79.-v, 72.25.-b, 85.75.-d}
\maketitle

Spintronic \citep{BaderParkinSpintronicsRevAnnRevCondensedMatterPhys2010,SpintronicsZutic} and magnonic devices \citep{YIGManonics,MagnonicsKruglyak,MagnonSpintronics} are receiving intense scientific attention due to their promise to deliver new technologies that can revolutionize computing and provide greater energy efficiency. In particular, tools for understanding phenomena such as angular momentum transfer across interfaces \citep{Tserkovnyak2002, Heinrich2011,Castel2012, Kajiwara, AdurDampingConfinedModeInYIG2014}, spin wave propagation in low dimensional and nanoscale systems \citep{demokritov2009spin,uBLSSwipWavesNanoscale}, domain wall motion \citep{ParkinMagDWMotion2003,FastDWMotion,AllWoodDomainWallLogic}, microwave-assisted switching \citep{ThirionMagSwitchingOrigPaper2003}, and relaxation and damping in small structures \citep{KrivoruchkoSpinWavesNanometer} are needed.  There is current interest in materials with more novel magnetic textures than simple ferromagnets, such as skyrmions \citep{NagosaTokura}. Electrical detection has been widely used for studying domain wall motion, but does not have imaging capabilities.  Optical techniques such as Brillouin light scattering (BLS) \citep{uBLSSwipWavesNanoscale} and the magneto-optic Kerr effect (MOKE)\citep{BaderMOKE} are also widely used but are ultimately limited by the optical diffraction limit. Scanned probe techniques can provide high spatial resolution but can be perturbative and may require a more challenging set-up such as vacuum and cryogenic environment to achieve high sensitivity. 

Nitrogen-vacancy (NV) centers in diamond have emerged as an attractive tool to study magnetic phenomena at the nanoscale, and they offer a way to convert magnonic signals into optical signals. NV centers offer a powerful magnetometry tool due to a potent combination of optical and magnetic properties that make the intensity of their photoluminescence (PL) dependent on their spin state. This has allowed detection of just a few resonant nuclear spins and nuclear magnetic resonance imaging with resolutions of tens of nanometers, all under ambient conditions and at room temperature \citep{Trio1DeVience2015,trio2Haberle, trio3Rugar}. NV centers have also been used to study domain wall hopping \citep{TetienneJacquesNVDomainWall}, the helical phase in FeGe \citep{DussauxDegenObservationOfLocalMagnetizationDynamicsInTheHelimagnetFeGe2015arXiv}, and spinwave modes in permalloy \citep{vanDerSarPyNVdiamondSpinwaves}.  High sensitivity to detect dynamic fields has been achieved by finding optimal NV centers with long lifetimes and manipulating them (and sometimes the target spins) with intricate microwave and optical pulse sequences. 

We have recently demonstrated a new approach \citep{WolfeNVFMR} to detect ferromagnetic resonance (FMR) using NV centers in nanodiamonds, whose short spin lifetimes and varied NV-center orientations typically render them unsuitable for conventional optically detected magnetic resonance (ODMR) based magnetometry. In contrast to all other reported approaches to detecting non-NV magnetic resonance signals, our technique requires no excitation at the NV frequency.  Rather, the intensity of NV PL responds directly to magnetic resonance excitation of the system under study, without the need for intricate pulsed magnetic resonance schemes. Here we report the extension of this approach to include the detection of several spinwave branches, both dipolar ($kd < 1$, where \emph{k} is the wavenumber and \emph{d} is the magnetic film thickness) and  dipolar-exchange ($1 \leq kd < 25$), as well as domain-related dynamics. Spectral studies of such complex magnetization dynamics have not been previously reported using NV centers. We also show that this technique provides spatially resolved information by measuring the dependence of the spectra on the position of the laser spot on the magnetic film.

\begin{figure}
\center{\includegraphics[width= 1\linewidth]{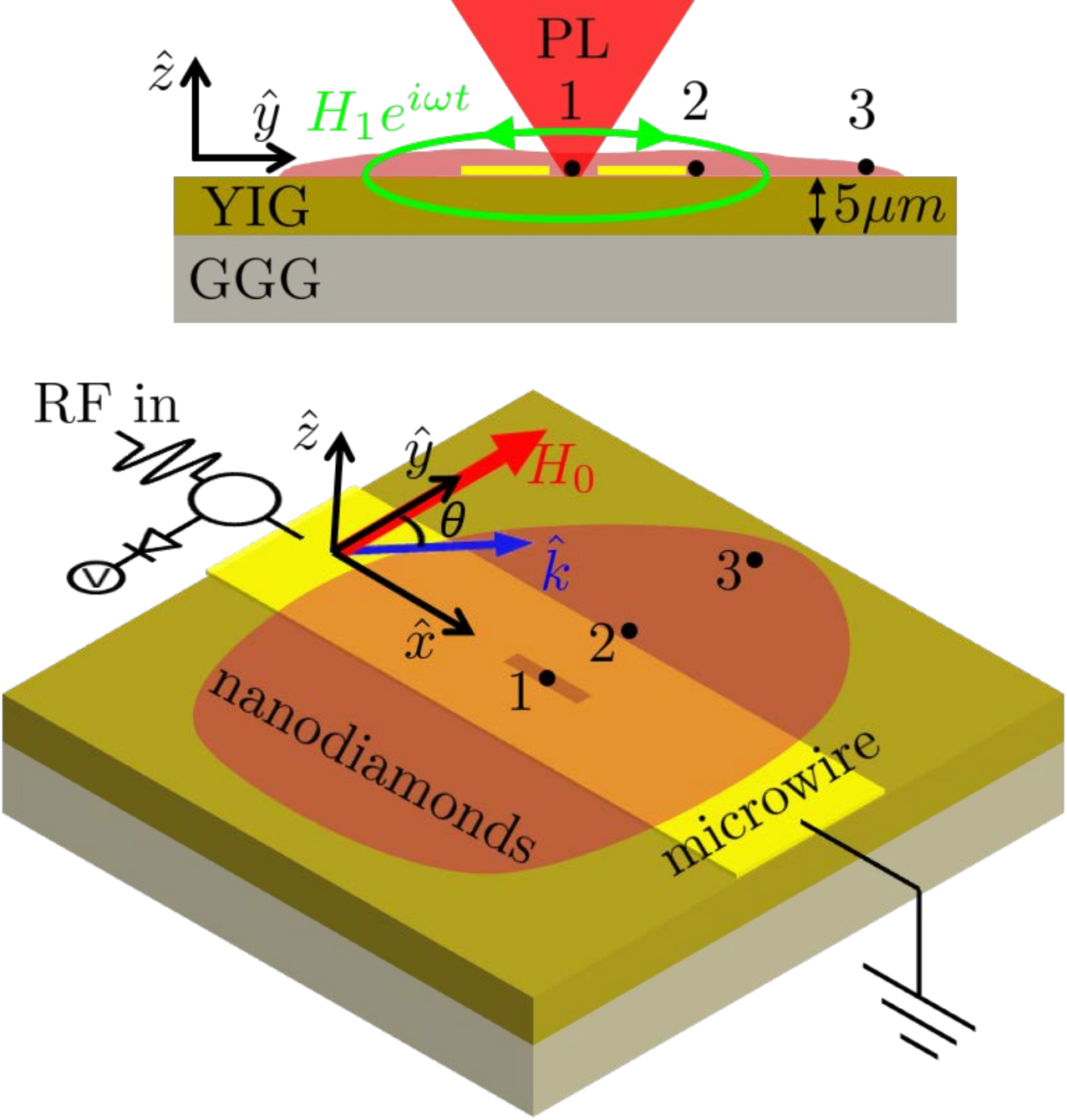}}
\caption{\textbf{Experimental schematic}: Experiments were performed on a 5\,$\mu$m thick YIG sample with a lithographically defined MW antenna. Nanodiamonds were dispersed on top and were in contact with the YIG. Changes in the luminescence of the NV centers in the nanodiamonds were recorded as a function of the static magnetic field, $H_0$ (large red arrow) and the frequency of the MW field, $H1$ (elliptical green arrow), at various locations on the sample.  Positions 1, 2, and 3 indicate the locations where NV signal was measured and correspond to panels (b), (c) and (d) of Fig.~\ref{fig:MainData}, respectively.   Magnetization dynamics in the ferromagnet were also monitored via the reflected MW power, which is shown in Fig.~\ref{fig:MainData} (e). This reflection signal spatially averages over the entire sample in contrast with the NV optical signal that measures local dynamics.}
	\label{fig:schematic}
\end{figure}

The sample is a 5\,$\mu$m thick film of yttrium iron garnet (YIG, Y$_3$Fe$_5$O$_{12}$) epitaxially grown on a (111) oriented gadolinium gallium garnet (GGG) substrate by liquid phase epitaxy.  A 400\,$\mu$m wide, 300 nm thick silver microstrip is lithographically patterned on top of the film as shown in Fig.\ref{fig:schematic} to apply microwaves (MW). A small window, 25\,$\mu$m X 200\,$\mu$m, is left open in the center of the silver wire to allow optical access to regions of the samples that experience various MW conditions. The shorted microstrip is driven by a MW generator. A film of nanodiamonds, 50-200 nm in size and containing up to a few thousand NV centers each, is dispersed on top of the sample.

\begin{figure}
\center{\includegraphics[width = 0.8\linewidth]{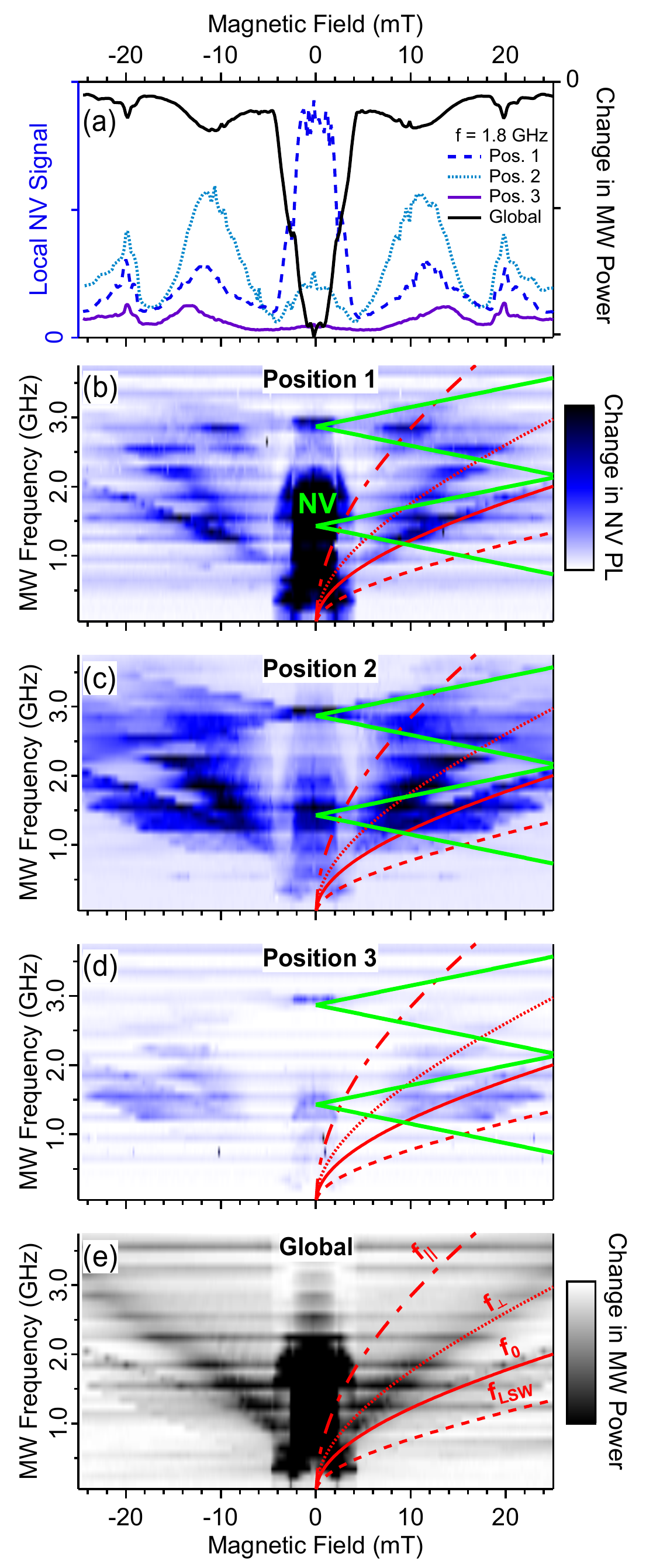}}
\caption{{\bf Spatially resolved, broadband spectroscopy of YIG using an ensemble of NV centers in nanodiamonds.}:  (a) Global change in reflected MW power (solid black line, right-hand axis) and local optical NV signal at three positions (blue shaded lined, left-hand axis) as a function of field $H_0$ at a MW frequency of 1.8 GHz.  (b, c, d) 2D maps of change in NV PL as a function of  field $H_0$ and frequency of $H_1$ for positions 1, 2, and 3 (see Fig. 1), respectively. The green lines indicate the outer limits of the NV ground (upper two lines) and excited state (bottom two lines) magnetic resonances due to the powder spectrum of NVs. (e) 2D map of the change in reflected MW power as a function of field $H_0$ and frequency of $H_1$. The various superposed red lines show the calculated dispersion relations for the various branches of the spinwaves for this YIG film.}
	\label{fig:MainData}
\end{figure}

PL is excited in the NV centers using a 520 nm laser, focused down to a $<2\,\mu$m spot, and is collected by a photodiode.  The NV-PL is recorded as a function of a static applied magnetic field $H_0$, applied in-plane and  perpendicular to the antenna (see Fig.~\ref{fig:schematic}), and the frequency, $f$, of the applied MW field $H_1$. We measure this PL signal at three different positions on the sample as indicated by labels 1, 2 and 3 in Fig.~\ref{fig:schematic}.  We measure a lock-in signal for both the PL and the reflected MW power by modulating the amplitude of $H_1$  at $\sim 1$ kHz. The fractional change of PL (lock-in voltage/DC level) is the primary data of interest and is presented in Fig.\ref{fig:MainData} (Note: NV lock-in signals are positive for decreases in NV PL). This is compared and contrasted to the change in reflected MW signal which also changes due to the power absorbed by FMR and other ferromagnetic dynamics. We emphasize that the reflected MW power is averaged over the entire sample while the NV-PL provides local information at the laser spot. 

Fig.\ref{fig:MainData}(a) shows representative spectra collected as $H_0$ is swept from -25 mT to +25 mT at 1.8 GHz MW excitation. Shown are the spectra collected by NV-PL at the three positions (left-hand axis) and the reflected MW power data (right-hand axis). We note that noise in NV-PL data (as given by the y-channel of the lock-in) is typically smaller than the thickness of the lines used in the graphs. This high signal-to-noise ratio was achieved with 500 ms lock-in time constant and no further averaging. 

A striking feature of this approach is the ability to detect a diverse array of excitations of the ferromagnet. Below, we describe these various spinwave excitations whose NV signal intensities are shown in Fig. \ref{fig:MainData} as a function of MW frequency and applied magnetic field. The intensities of these features change between the four spectra due to the varying microwave conditions across the sample. We will first describe these features and then discuss their spatial variation.

The peak at $H_0\sim 20$ mT in Fig.\ref{fig:MainData}(a) corresponds to the signal from  the spinwave modes defined by the sample thickness ($kd \approx 0$). These spinwaves propagate along the direction of the external field $H_0$ (see Fig.\ref{fig:schematic}). Later in the text we refer to them as  longitudinal spinwaves (LSW). These spinwaves form the higher shoulder of the peak at $H_0$ = 20 mT, reaching wavevectors as high  as $kd \approx 5$ with increasing $H_0$. The excitation of spinwaves with such relatively large wavevectors arises as a consequence of the spatial inhomogeneity of the MW field from the antenna as well as from the condition $\theta = 0$ on the angle between the propagation direction and the external field $H_0$. We associate the second peak (observed for $10 < |H_0| < 14$ mT) with parametrically excited spinwave modes by perpendicular pumping in the $H_{0} \perp H_{1}$ geometry (discussed below). These spinwaves have wavectors on the order of $kd \approx 25$ \citep{Wiese_RF_param_thresh_YIG_1994} and are oriented at angles $15^{\circ} <\theta < 45^{\circ}$ relative to the applied field $H_0$ . In fields $< 4$ mT we see the largest feature arising from the unsaturated multi-domain state of the YIG. Any changes to the PL due to direct NV spin resonance are weak, compared to those due to the ferromagnet, at 1.8 GHz for our polycrystalline nanodiamond powder.

The spinwave excitations discussed above are identified using the field-frequency sweeps of NV and microwave signals shown in Fig.~\ref{fig:MainData}. The panels (b)-(d) correspond to NV PL at positions 1, 2, and 3, and (e) to changes in reflected MW power. The LSW modes with $kd \approx 0$ are approximately identified by the Kittel FMR mode $f_0 = \gamma\sqrt{(H_0(H_0+4\pi M_s))}$. The higher field cutoff for these LSW modes is marked by $f_{LSW}$ (red  dashed line), which corresponds to $kd = 5$. This $f_{LSW}$ vs. $H_0$ dependence is calculated using the analytical solution from previous work\cite{Kalinikos_theory_SW_original}(see supplementary information). Thus, curves $f_0$ and $f_{LSW}$ define the range of spinwave wavevectors $0 \lesssim kd \lesssim 5$.

The curves $f_{\bot}$ (second peak in panel (a)) and $f_{\|}$ depict spinwaves generated via first order Suhl instability processes \citep{Gurevich_MOandW} where spinwaves are excited at half the frequency of the driving MW field, $H_1$. This process favors excitation of spinwaves with the lowest damping and group velocity \citep{Gurevich_MOandW}. The curve $f_{\bot}$ corresponds to the excitation of dipolar-exchange spinwaves by a perpendicular pumping field, $H_{0} \perp H_{1}$ with $kd \approx 25$\citep{Wiese_RF_param_thresh_YIG_1994} and $15^{\circ} <\theta < 45^{\circ}$. The red dot-dashed curve, labeled $f_{\|}$, shows the dynamics driven by parallel pumping \citep{Gurevich_MOandW},  $H_1||H_0$, where effective excitation takes place for spinwaves with wavevectors on the order of $kd \sim 0$ and $5\ (\theta = 90^{\circ})$\citep{Wettling_BLS_parallel,Kabos_ParalPump_BLS_1997}. At the given microwave powers ($\sim25$ mW going into the sample) the spinwaves can exist in a broad range of wavevectors and propagation angles, $\theta$ \citep{Wettling_BLS_parallel,Kabos_ParalPump_BLS_1997,Kabos_BLS_perp_pump_high_power_1994}. Moreover, 3- and 4-magnon scattering further increases the range of excited spinwaves\cite{Patton_3m_2009,Kabos_second_instab_1996}. Our data, however, do not allow us to separate the roles each of these processes play in affecting the observed PL signal.

The signal arising from dynamics in the multi-domain state of the YIG are seen for $H_0 < 4$ mT. This signal most likely originates from the resonance modes of individual domains, \citep{Spin_Dyn_conf_str} as well as dipolar spinwaves, which have complicated spectra \citep{Vashnoskij_SW_domains_1996,Vashnoskij_RUS_SW_domains_1998}.

The data in Fig.~\ref{fig:MainData} demonstrate the coupling between NV centers and the ferromagnet over a broad range of fields and frequencies. A key point is that the native resonances of the NV need not overlap signals arising from the ferromagnet. The NV resonances from the randomly oriented diamonds fall between the extrema of the powder pattern which are shown by green lines.

The data in Fig.\ref{fig:MainData}(b-d) also show a pronounced sensitivity of the NV signal to the position of the laser spot on the sample, and clearly contrast with the global reflected MW power data.  The NV signal provides information about local magnetic dynamics, as induced by the local microwave field (and the uniform static magnetic field) within the area illuminated by the laser. 

Our experimental arrangement makes possible the excitation of the different dynamic modes of the YIG. At position 1, as shown in Fig. 1, the MW field, $H_1$, is nearly parallel to the external biasing field, $H_0$, and thus induces only a small torque on the YIG magnetization. Therefore, the efficiency of linear excitation is low. But these are ideal conditions for parallel pumping via a first order Suhl instability; see curve $f_{\|}$ in Fig.\ref{fig:MainData}(b). At position 2 the MW fields are stronger (as evidenced by the larger NV ground state resonance) and perpendicular to the film surface and hence to the magnetization; ideal conditions for linear drive of spinwaves, which are generated by the strongly non-uniform MW fields at the edge of the microstrip (curves $f_0$ and $f_{LSW}$). However, at high MW power levels the excitation of spinwave via a first order Suhl order instability is dominant for this position, curve $f_{\bot}$ in Fig.\ref{fig:MainData}(c).

Finally, at position 3, while most the features are detectable, they are much smaller due to the much weaker MW fields given the separation from the microstrip. The signal here most likely results from a combination of direct excitation by the MWs and by the propagating spinwaves excited at the edges of the microstrip (position 2).  The spinwave band, $f_{\bot}$, at position 3 seems to be shifted to higher field (see Fig.~\ref{fig:MainData}(a)) and suggests propagating spinwaves.  The field is shifted because the spinwaves at the center of the band have small group velocity, while the ones at higher field have a larger group velocity and can thus travel the 500\,$\mu$m to the location of the laser spot where they are detected.

The biggest difference between position 1 and 2 is in the unsaturated state.  Position 1, in the center of the wire, shows the strongest NV signal from the multi-domain state and matches closely to the global reflected MW signal because the reflected MW signal will be dominated by the region under the wire.  This is in contrast to position 2 where the NV signal from the spinwave band and even the uniform mode is larger than the domain state signal. The intensity of the microwaves at position 2 are stronger than at position 1 and cannot explain the relative change of the domain-state signal between the positions. This is most likely due to the  out-of-plane tilted domains \citep{Xia_OOPDomains} that arise from growth-induced anisotropy in these YIG films. The out-of-plane tilted domains are much more effectively excited by the in-plane microwave field that exists at position 1 (which, being perpendicular to the magnetization can provide a greater torque). Given the significant interest in the various technological uses of domains and domain walls, it should be investigated in further detail. NV detection may be an ideal way to study dynamics in domains because of the large response and spatial resolution.

\begin{figure}
\center{\includegraphics[width=1\linewidth]{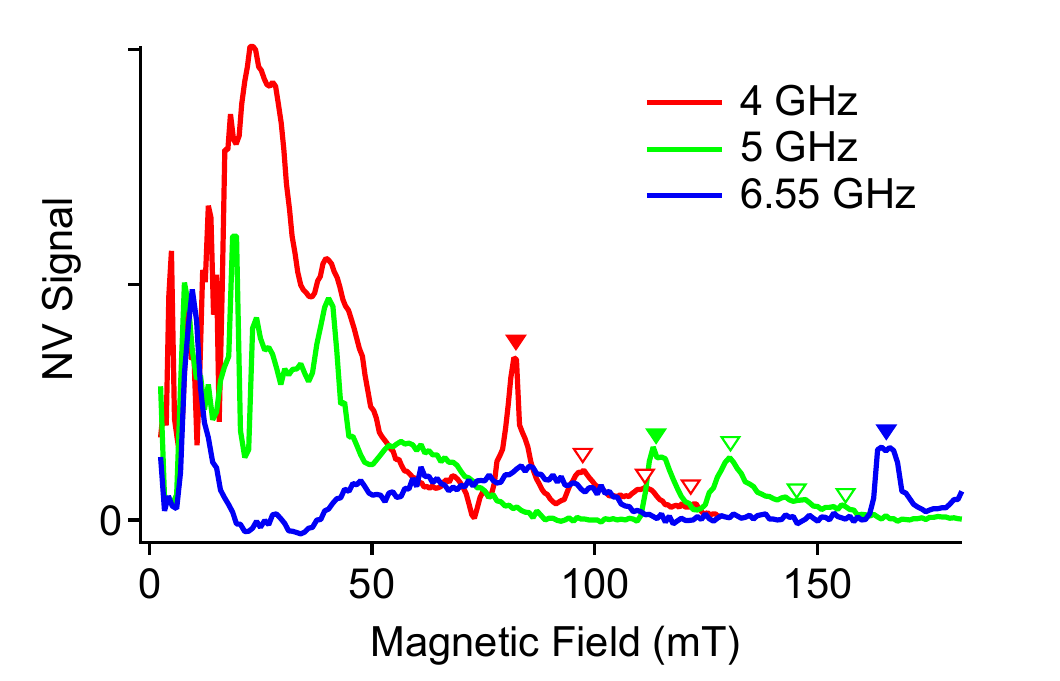}}
\caption{Measurement of ferromagnetic dynamics using NV centers in magnetic fields exceeding those typically used for conventional ODMR magnetometry. The solid triangles correspond to uniform FMR at the given frequency. The open triangles correspond to spinwave modes due to the inhomogeneity of microwaves resulting from the window in the antenna structure. The modes have the same $kd = 0.95, 2.2$ and $4$ for all three frequencies, with higher $kd$ corresponding to higher magnetic field.}
	\label{fig:HighFreqData}
\end{figure}

The NV response is sensitive to ferromagnetic dynamics up to fields and frequencies higher than shown in Fig.\ref{fig:MainData}.  The evolution of the ferromagnetic spectrum up to 6.55 GHz and 180 mT, the limit of our current set-up, is shown in Fig. \ref{fig:HighFreqData}.  The uniform FMR is indicated by the solid triangular symbols for each frequency. At 6.55 GHz (blue line) we see the uniform mode at 165 mT.  This field is higher than reported for any magnetometry experiments with nitrogen vacancy centers thus far. Measuring NV signals at higher magnetic field with conventional ODMR can be challenging due to the decay of ODMR contrast with field and due to the ill effects of fields transverse to the NV axis \citep{LukinNVTransverseField,Transverse}, which can exist either due to non-ideality of the experimental alignment or limitations of the measurement geometry. These are especially true in the case of nanodiamonds, where pulsed ODMR is challenging for ensembles even at modest fields.  This highlights the potential versatility of our "off-resonant" modality of detection for the study of magnetic dynamics over a broad range of fields and frequencies.

We also see higher field peaks (relative to FMR and indicated by open triangles in Fig.\ref{fig:HighFreqData}), which correspond to the LSWs ($\theta = 0$) and are generated by the small, non-uniform perpendicular field at the edges of the 25$\,\mu$m window in the miscostrip line (see Fig.\ref{fig:schematic}). The wavevectors \emph{k} corresponding to these excitations is approximately given by the standing-wave condition determined by the width of the window $W$, $k \approx n\pi/W$, where $n$ is an odd number. The wavevectors are better calculated using formulas from Kalinikos\citep{Kalinikos_theory_SW_original}(see supplementary information). The peaks correspond to $kd = 0.95, 2.2 $ and $4$, with $kd$ increasing with magnetic field.
	
To conclude, we have shown that NV centers are sensitive to magnetic dynamics over a broad range of magnetic fields and microwave frequencies, and that the NV centers provide localized sensitivity to magnetic dynamics in the immediate vicinity of the NV detector. These results, combined with the atomic nature of the NV defects, indicate that NV centers could be used in this "off-resonant" modality to study ferromagnetic dynamics on the nanoscale and in novel magnetic textures. Domain wall motion is currently a field of intense interest and technological promise \citep{ParkinMagDMMem,TetienneJacquesNVDomainWall}.  The fact that we see such strong signals from the dynamics of this state presents a promising potential avenue for optical read out of domain wall motion and domain dynamics. These results can enhance a recent proposal to use FMR of a ferromagnetic element for amplifying the magnetic resonance signal from a nearby target nuclear spins \citep{YacobyLoss2015}, and highlight the potential for using spinwave modes in addition to the uniform mode for such a scheme. 

Funding for this research was provided by the ARO through award number W911NF-12-1-0587 (NV diamond optical detection and measurement of magnetic signals), the Center for Emergent Materials at the Ohio State University, an NSF MRSEC through award Number DMR-0820414 (spin wave dynamics, sample growth and characterization).

\bibliography{SpatiallyResolvedDetectionOfComplexFerromagneticDynamicsUsingNV_v15}
\end{document}


\title{Supplementary material: Spatially resolved detection of complex ferromagnetic dynamics using optically detected NV spins}

\author{C. S. Wolfe}
\affiliation{Department of Physics, The Ohio State University, Columbus, Ohio 43210, USA}

\author{S. A. Manuilov}
\affiliation{Department of Physics, The Ohio State University, Columbus, Ohio 43210, USA}

\author{C. M. Purser}
\affiliation{Department of Physics, The Ohio State University, Columbus, Ohio 43210, USA}

\author{R. Teeling-Smith}
\affiliation{Department of Physics, The Ohio State University, Columbus, Ohio 43210, USA}

\author{C. Dubs}
\affiliation{INNOVENT e.V. Technologieentwicklung, Pruessingstrasse 27 B, 07745 Jena, Germany}

\author{P. C. Hammel}\email{hammel@physics.osu.edu*}
\affiliation{Department of Physics, The Ohio State University, Columbus, Ohio 43210, USA}

\author{V. P. Bhallamudi}\email{bhallamudi.1@osu.edu*}
\affiliation{Department of Physics, The Ohio State University, Columbus, Ohio 43210, USA}


\date{\today}


\maketitle

\section{Spinwave dispersion}
The spinwave dispersion relation developed in Ref.\citep{Kalinikos_theory_SW_original} provides a simple analytical formulation for different material parameters and applied magnetic fields. It  takes into account both dipolar and exchange interactions.  For the longitudinal spinwaves (LSW) with $k \parallel H_{0}$ we have
\begin{equation}
\label{eq1}
	\omega/2\pi = \sqrt{\Omega_k\left[ \Omega_k + 4\pi M_s\gamma(1-P_k)\right]}
\end{equation}
where wavevector dependent exchange $\Omega_k$ and dipolar $P_k$ terms are given in Ref.\citep{Kalinikos_theory_SW_original} by Eqs. (16) and (19), respectively. The gyromagnetic ratio is $\gamma = 2.8$ MHz/Oe and saturation magnetization is $4\pi M_s = 1760$ G for our $d=5\ \mu$m thick YIG film. 

In our calculations we assume the surface spins are free (unpinned spins). This is a valid assumption for YIG films grown by liquid phase epitaxy \citep{WIGEN}. 

\begin{figure}[h]
\center{\adjincludegraphics[width=1\linewidth ,trim= {0\width} {0.2\height} {0\width}  {0.21\height},clip]{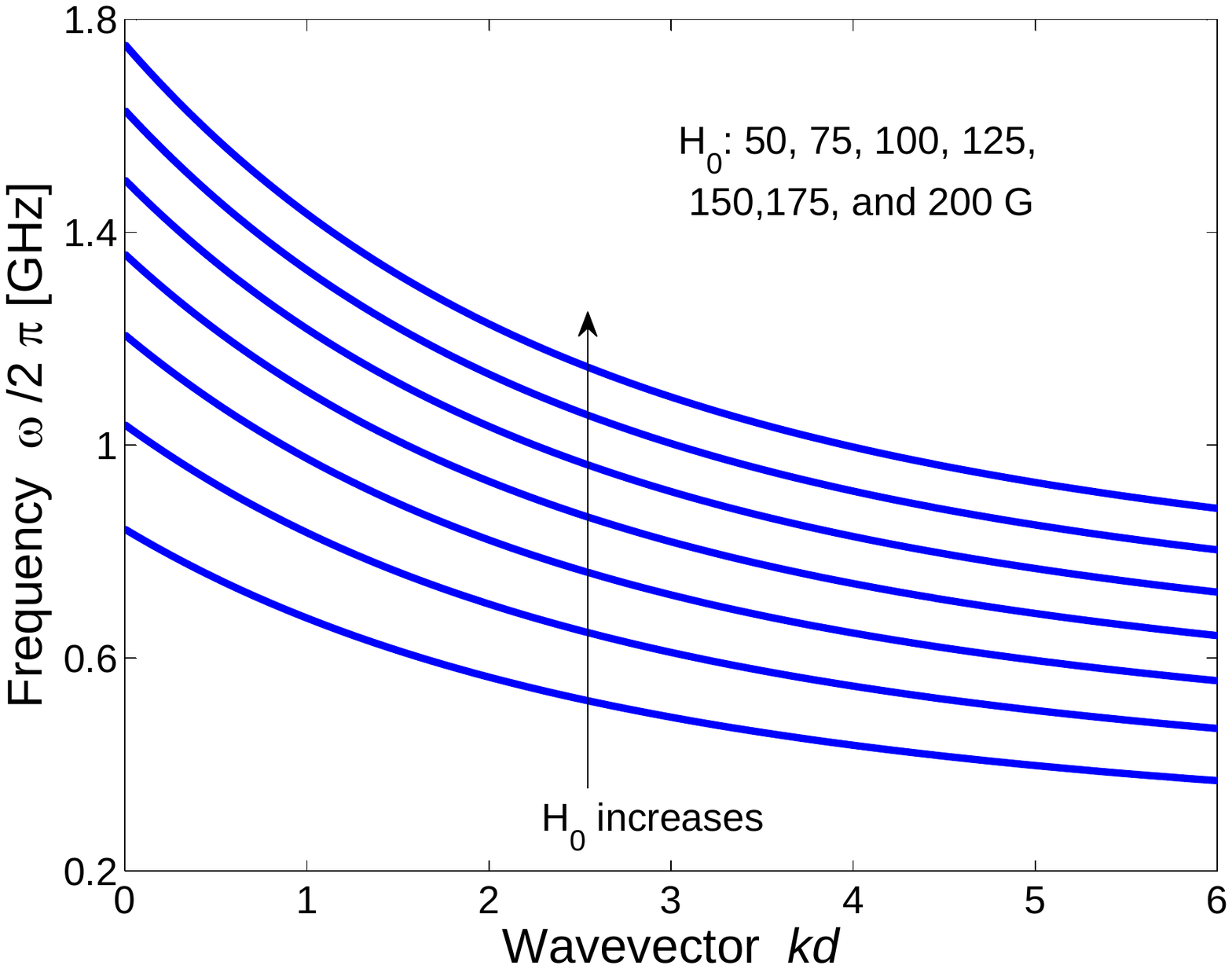}}
\caption{Dispersion curves for $k \parallel H_{0}$ spinwaves at different applied fields $H_0$.}
	\label{fig:SW_disp_exmpl}
\end{figure}

Figure Fig.\ref{fig:SW_disp_exmpl} shows a family of the spinwave dispersion curves calculated using Eq.\ref{eq1} for different values of the applied field $H_0$. Similar calculations have been performed in order to estimate wavevector numbers $kd$ using experimental data from Fig.2 (curve $f_{LSW}$) and Fig.3 in the main text.

\bibliography{suppl_matNotes}